\documentclass[twocolumn,aps]{revtex4-1}
\usepackage{bm}
\usepackage{amsmath}
\usepackage{epsfig}
\usepackage{graphicx}
\usepackage{subfigure}

\def\gazz{\mathrel{\mathchoice {\vcenter{\offinterlineskip\halign{\hfil
$\displaystyle##$\hfil\cr>\cr\sim\cr}}}
{\vcenter{\offinterlineskip\halign{\hfil$\textstyle##$\hfil\cr>\cr\sim\cr}}}
{\vcenter{\offinterlineskip\halign{
\hfil$\scriptstyle##$\hfil\cr>\cr\sim\cr}}}
{\vcenter{\offinterlineskip\halign{\hfil$\scriptscriptstyle##
$\hfil\cr>\cr\sim\cr}}}}}		

\def\be{\begin{equation}}
\def\lan{\left\langle}
\def\ran{\right\rangle}
\def\ee{\end{equation}}
\def\barr{\begin{array}}
\def\earr{\end{array}}

\def\l{\left}
\def\r{\right}
\def\dis{\displaystyle}
\def\ed{\end{document}}

\def\blamf{\lambda_{F_k}}

\def\bhh{{\mbox{\boldmath $H$}}}

\def\cs{{\bf s}}
\def\cf{{\bf f}}

\def\spin{\frac{1}{2}}

\begin{document}

\title{Localization-Delocalization Transitions in Bosonic Random Matrix Ensembles}
\author{N. D. Chavda$^1$ and V. K. B. Kota$^{2}$}

\address{$^1$Department of Applied Physics, Faculty of Technology and
Engineering, Maharaja Sayajirao University of Baroda, Vadodara 390 001, India\\
$^2$Physical Research Laboratory, Ahmedabad 380 009, India}

\begin{abstract}

Localization to delocalization transitions in eigenfunctions are studied for
finite interacting boson systems by employing one- plus two-body embedded
Gaussian orthogonal ensemble of random matrices [EGOE(1+2)]. In the first
analysis, considered are bosonic EGOE(1+2) for two-species  boson systems with a
fictitious ($F$) spin degree of freedom [called BEGOE(1+2)-$F$]. Numerical
calculations are carried out as a function of the two-body interaction strength ($\lambda$). It is shown that, in the region (defined by
$\lambda>\lambda_c$) after the onset of Poisson to GOE transition in energy
levels, the strength functions exhibit Breit-Wigner to Gaussian transition for
$\lambda>\blamf>\lambda_c$. Further, analyzing information entropy and
participation ratio, it is established that there is a region defined by
$\lambda\sim\lambda_t$ where the system exhibits thermalization. The $F$-spin
dependence of the transition markers $\blamf$ and $\lambda_t$ follow from the
propagator for the spectral variances. These results, well tested near the center of the spectrum and extend to the region within
$\pm2\sigma$ to $\pm3\sigma$ from the center ($\sigma^2$ is the spectral variance), establish universality of the
transitions generated by embedded ensembles. In the second analysis,
entanglement entropy is studied for spin-less BEGOE(1+2) ensemble and shown that the
results generated are close to the recently reported results
for a Bose-Hubbard model.

\end{abstract}

\maketitle
\section{Introduction:}

Transition from localization to delocalization phases in many-body quantum
systems, due to inter particle interaction, has received a great attention in
recent years
\cite{Znid08,Pal-10,Khatami2012,JHB2012,RNand2015,Huse2007,Altman2015}. Now, a
connection between the physics of the localization-delocalization transition and
the physics of thermalization is well
established~\cite{Sengupta,Rigol16,BISZ2016}. This is possible mainly due to
major developments in the experimental study of many-particle quantum systems
such as ultra-cold gases trapped in optical
lattices~\cite{Kondov2015,Schreiber2015,Bordia2015} and ions~\cite{Smith2015}.
It has been established that many-body interacting systems in which localization
occurs, a phenomenon termed many-body localization (MBL), avoid thermalization
due to the emergence of extensively many local integrals of motion. On the other
hand, non-integrable quantum systems thermalize and the eigenstate
thermalization hypothesis (ETH) is considered to be the underlying mechanism for
thermalization~\cite{ETH, Rigol2008}. ETH essentially means that the expectation
values of typical observables in the eigenstates of a many-body interacting
system follow ergodicity principle. Various aspects related to the role of
localization and chaos, statistical relaxation, eigenstate thermalization and
ergodicity principle have been studied by many groups using lattice models of
interacting spins (fermionic as well as bosonic systems) \cite{BISZ2016,Lea-12, Kollath,
ETH-1, ETH-2, Iz-PRL, Manan}.

Embedded Gaussian Orthogonal Ensemble (EGOE) of random matrices (for
time-reversal and rotationally invariant systems) with one plus two-body
interactions for fermions [called EGOE(1+2)], introduced in the past~
\cite{MF,Brody,FI1,FI2,vkb2001}, are paradigmatic models to study the
dynamical transition from integrability to chaos in isolated finite interacting
many-body quantum systems \cite{vkb2001,PW-07,Ko-14,BISZ2016}. It is important to note
that EGOE are generic, though analytically difficult to deal with, compared to
the lattice spin models as the latter are associated with spatial coordinates
(nearest and next-nearest neighbor interactions) only. Moreover, it is seen that
the universal properties derived using EGOE apply to systems represented by
lattice spin models. EGOE(1+2), as a function of the two-body interaction
strength $\lambda$ (measured in units of the average spacing between the
one-body mean-field single particle  levels), exhibits three transition or chaos
markers $(\lambda_c,\blamf,\lambda_t)$: (a)~as the two-body interaction is
turned on, level fluctuations exhibit a transition from Poisson to GOE at
$\lambda=\lambda_c$; (b) with further increase in $\lambda$, the strength
functions (also known as the local density of states) make a transition from
Breit-Wigner (BW) form to Gaussian form at $\lambda=\blamf > \lambda_c$; and (c)
beyond $\lambda=\blamf$, there is a region of thermalization around
$\lambda=\lambda_t$ where basis dependent thermodynamic quantities like entropy
behave alike. For further details of the three chaos markers for EGOE(1+2) for
spin-less fermion systems, see~\cite{Ko-14, KoReta}. It was also established
that interacting fermion systems with fermions carrying spin ($\cs = 1/2$)
degree of freedom giving EGOE(1+2)-$\cs$ also exhibit these three transition
markers and their spin dependence was also deduced~\cite{Ma-PRE}; the propagator
of the spectral variances explains the spin dependence. It is
important to add that the transitions mentioned above are inferred from large
number of numerical calculations and they are well verified to be valid around  the  center of the spectrum and extend in to the region within $\pm 2\sigma$ to $\pm 3\sigma$ from the center ($\sigma$ is the spectral width), i.e. in the  bulk part of the spectrum. Also, the chaos markers are  in general energy dependent; see Section~\ref{sec3} for an example. One observes
deviations in the spectrum ends and this may be due to the the fact that the eigenvalue density
generated by EGOE(1+2) is only asymptotically a Gaussian~\cite{MF,Ko-14}.

For bosonic systems, embedded Gaussian orthogonal ensembles of one- plus
two-body interactions [denoted by BEGOE(1+2)] for finite isolated interacting
spin-less many-boson systems have been analyzed in detail
\cite{Pa-00,Ag-01,Ag-02,Ch-0304,Ch-PLA}, as they are generic models for finite
isolated interacting many-boson systems. For $m$ bosons in $N$ single particle
(sp) states, in addition to the dilute limit (defined by $m \rightarrow \infty$,
$N \rightarrow \infty$, $m/N \rightarrow 0$), another limiting situation, namely
the dense limit (defined by $m \rightarrow \infty$, $N \rightarrow \infty$, $m/N
\rightarrow \infty$) is also possible. Such a limiting situation is absent for
fermion systems. Hence the focus was on the dense limit in BEGOE investigations
\cite{Pa-00,Ag-01,Ag-02,Ch-0304,Ch-PLA}. In the strong interaction limit,
two-body part of the interaction dominates over one-body part and therefore
BEGOE(1+2) reduces to BEGOE(2). In the dense limit, the eigenvalue density takes
Gaussian form irrespective of the strength ($\lambda$) of the two-body
interaction for these ensembles \cite{Pa-00,Ag-02,KP-80}.  Just as spin-less
fermionic EGOE(1+2), the BEGOE(1+2) also exhibits, as $\lambda$ increases, three
chaos markers $(\lambda_c,\blamf,\lambda_t)$. See \cite{Ch-0304,Ch-PLA,Ko-14}
for further details.

Going beyond spin-less boson systems, very recently BEGOE for two species boson
systems with a fictitious $F$ spin-1/2 degree of freedom [called BEGOE(1+2)-$F$]
\cite{Ma-F} and for a system of interacting bosons carrying spin-one degree of
freedom [called BEGOE(1+2)-$S1$] \cite{Deota-S1} are introduced and their
spectral properties are analyzed in detail. Here, it is important to note that
the $F$-spin for bosons is similar to the $F$-spin in the proton-neutron
interacting boson model  ({\it pn}IBM) of atomic nuclei  \cite{Ia-87,Ca-05}.
Similarly, BEGOE(1+2)-$S1$ is an important extension to apply BEGOE to spinor
BEC discussed in \cite{Pe-10,PRA}. The purpose of the present paper is firstly
to establish universality of the localization-delocalization transitions
generated by embedded ensembles by showing that BEGOE(1+2)-$F$ generates
$\blamf$ and $\lambda_t$ transition markers (in addition to the $\lambda_c$
marker studied in \cite{Ma-F}) just as in the fermionic ensembles, without and
with spin, and bosonic ensembles without spin. Secondly, going beyond the
previous analysis of embedded ensembles, here for the first time, the bipartite
entanglement entropy is studied using embedded ensemble for spin-less boson
systems and shown that the results are close to those obtained recently using
Bose-Hubbard models.

The paper is organized as follows. In Section~\ref{sec2}, briefly introduced is
the BEGOE(1+2)-$F$ model. Section~\ref{sec3} is on the $\blamf$ marker and gives
the results for BW to Gaussian transition in strength functions. Results for the
participation ratio (PR) and the information entropy ($S^{\mbox{info}}$) are
described in Section~\ref{sec4}. Section~\ref{sec5} deals with the
thermalization $\lambda_t$ marker using information entropy. In
Section~\ref{sec6}, the bipartite entanglement entropy ($S^{EE}$) is studied and
correlations of it with PR are presented. Finally, Section~\ref{sec7} gives some
concluding remarks. In Appendix A, for easy reference, collected are the
definitions of strength functions, PR and $S^{\mbox{info}}$ and also some EGOE
formulas that are used in the analysis presented in Sections
\ref{sec3}--\ref{sec5}.

\section{Embedded bosonic ensembles with $F$-spin:}
\label{sec2}

\subsection{Definition and construction:}
\label{subsec:1}

Consider a system with $m$ ($m >2$) bosons distributed in $\Omega$ number of sp
orbitals each with spin $\cf =\spin$. With $\Omega$ number of orbital degrees of
freedom and two spin ($m_\cf=\pm \spin$) degrees of freedom, the total number of
sp states is $N=2\Omega$. The sp states for bosons are denoted by $\l.\l|
i,m_\cf=\pm \spin\r.\ran_s$ with $i=1,2,\ldots,\Omega$ and the two particle
symmetric states are denoted by $\l.\l|(ij)f,m_f\r.\ran$ with the two particle
spin $f=0$ or $1$. For one plus two-body Hamiltonians preserving $m$ particle
spin $F$, the one-body Hamiltonian is $h(1) = \sum_i \epsilon_i \; n_i$, where
the orbits $i$ are doubly degenerate, $n_i$ are number operators  and
$\epsilon_i$ are sp energies. Similarly the two-body Hamiltonian $V(2)$ is
defined by the two-body matrix elements $V^f_{ijkl}={_s}\lan (kl)f,m_f \mid V(2)
\mid (ij)f,m_f\ran_s$ and they are independent of the $m_f$ quantum number.
Thus, $V(2)=V^{f=0}(2) + V^{f=1}(2)$; the sum here is a direct sum with
dimensions $\Omega(\Omega-1)/2$ and $\Omega(\Omega+1)/2$ respectively.  Then
BEGOE(1+2)-$F$ is defined by the Hamiltonian $H$ that preserves total $m$
particle spin $F$,

\be
H= h(1) + \lambda V(2) \;.
\label{eq-1}
\ee
Here, $\lambda$ is the strength of the two-body interaction $V(2)$. In
principle, it is possible to use a more general $H$ with the strengths of the
$V^{f=0}(2)$ and $V^{f=1}(2)$ to be different but this is not considered in this
paper. It is important to note that the $m$ particle states can be classified
according to the algebra $U(N) \supset U(\Omega) \otimes SU(2)$ with $SU(2)$
generating spin $F$. As $H$ preserves $F$, it is a scalar in spin $SU(2)$ space.
This recognition allows one to derive formulas for the spectral variances
discussed in Sec.~\ref{subsec:2}.

BEGOE(1+2)-$F$ is generated by the action of $H$ on many-particle basis states
with $V^{f=0}(2)$ and $V^{f=1}(2)$ being independent GOEs in two-particle
spaces; variances of the matrix elements in both GOEs are taken to be unity.
Given the sp energies and the two-body matrix elements, the many-particle
Hamiltonian $H$ is first constructed in the $M_F = M_F^{min}$ [$M_F^{min} =
0(\spin)$ for $m$ even(odd)] representation using the spin-less BEGOE formalism
and then states with a given $F$ are projected using the spin projection
operator $F^2$; see \cite{Ma-PRE, Ma-F} for details. Thus, the many-particle
Hamiltonian $H$ will be a block diagonal matrix with each diagonal block
corresponding to a given spin $F$ \cite{Ma-F}.

Earlier, by studying the distribution $P(r)$ of the ratio of consecutive level
spacings, that do not require spectral unfolding and introduced by Oganesyan and
Huse \cite{Huse2007}, to investigate many-body localization
\cite{Huse2007,OPH2009}, it was demonstrated that the embedded ensembles
including BEGOE(1+2)-$F$ exhibit GOE level fluctuations for sufficiently strong
interaction strength $\lambda$ \cite{CK2013,CHK2014}. Note that Atas et al.
\cite{ABGR-2013} derived expressions for the $P(r)$ distribution for the
classical GOE, GUE and GSE ensembles of random matrices. In addition, using the
nearest neighbor spacing distribution (NNSD), it was shown that BEGOE(1+2)-$F$
exhibits Poisson to GOE transition as the interaction strength $\lambda$ is
increased and the transition marker $\lambda_c$ is found to decrease with
increasing $F$-spin. Moreover, the ensemble-averaged spectral variance explains
the decrease in $\lambda_c$ with increasing $F$  \cite{Ma-F}. With $\Omega=4$,
$m=10$, the $\lambda_c$ values for these ensembles are found to be $0.039$,
$0.0315$ and $0.0275$ for $F=0$, $2$ and $5$ respectively. Going beyond these
investigations where only eigenvalues were analyzed, in the present study, our
focus is on eigenfunctions. Numerical calculations are carried out using fixed
set of sp energies $\epsilon_i=i+1/i$ as in our previous studies \cite{Ma-PRE,
Ma-F} and generated 100 member BEGOE(1+2)-$F$ ensembles with $\Omega=4$ and
$m=10$. The $H$ matrix dimensions for various $F$ values are $d(m,\Omega,F) =
196, 540, 750, 770, 594$ and $286$ for $F = 0-5$; $\sum_{F=F_{min}}^{F_{max}}
(2F+1) \; d(m,\Omega,F) = {{2\Omega+m-1}\choose{m}}$.

\subsection{BEGOE(1+2)-$F$ spectral variances and their spin dependence:}
\label{subsec:2}

Here, we will briefly discuss the structure of the ensemble averaged spectral
variance for BEGOE(1+2)-$F$ as they will provide an analytical understanding of
the spin dependence of the chaos markers. With fixed $h(1)$, the ensemble
averaged spectral variance  $\overline{\sigma^2_H(m,F)}$ follows simply from
Eq.~\eqref{eq-1},

\be
\overline{\sigma^2_H(m,F)}= \sigma^2_{h(1)}(m,F) +\lambda^2\;
\overline{\sigma^2_{V(2)}(m,F)}\; \;.
\label{eq.sigmaH}
\ee
Note that `overline' denotes ensemble average. The $\sigma^2_{h(1)}(m,F)$ and
$\sigma^2_{V(2)}(m,S)$ are the variances generated by the $h(1)$ part and the
perturbation $V(2)$ respectively. Now, for a uniform sp spectrum having unit
average level spacing, the $\sigma^2_{h(1)}(m,F)$ is  given as \cite{Ma-F},

\be
\label{eq.sig2h1}
\sigma^2_{h(1)}(m,F) = \frac{1}{12}\Big[m(\Omega-2)(\Omega+m/2) +
2\Omega F(F+1) \Big]\;.
\ee
Similarly, the ensemble averaged variance generated by the two-body interaction
$V(2)$ is given as $\overline{\sigma^2_{V(2)}(m,F)} = Q(\Omega, m, F)$. Explicit
formula, derived using trace propagation methods, for the variance propagator
$Q(\Omega, m, F)$ is given in Appendix-A and it is taken from
\cite{Ma-F,Ko-14}.  The propagator $Q(\Omega, m, F)$ (see  Fig.~6 in
\cite{Ma-F}), for bosonic systems, increases with increasing spin $F$. This
trend is opposite for fermionic systems \cite{Ma-PRE}.

\section{Transition in strength functions: $\blamf$ marker}
\label{sec3}

\begin{figure}[t]
\begin{center}
\includegraphics[width=\linewidth]{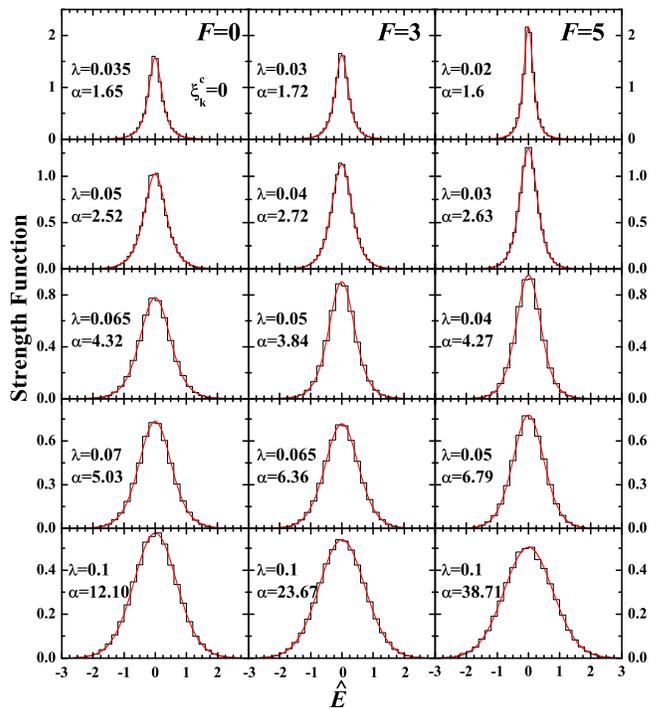}
\end{center}
\caption{Strength functions for $k$ states averaged over
$\hat{\xi^c_k} - \Delta_k \leq \hat{\xi_k} \leq \hat{\xi^c_k} + \Delta_k$ with
$\hat{\xi^c_k} =0$. Strength functions are shown for five different values of
$\lambda$ and they are for a 100 member BEGOE(1+2)-$F$ ensemble. Calculations
are for $\Omega=4$, $m = 10$ system with spins $F = 0, 3$ and $5$. Note that the
width $\sigma_{F_k(m,F)}$ of the strength functions is different from the
spectral width $\sigma_H(m,F)$. Continuous curves in the figures correspond to
the $t$-distribution given by Eq.~(\ref{fkstu}). See text for further details.}
\label{fke}
\end{figure}

\begin{figure}[t]
\begin{center}
\includegraphics[width=\linewidth]{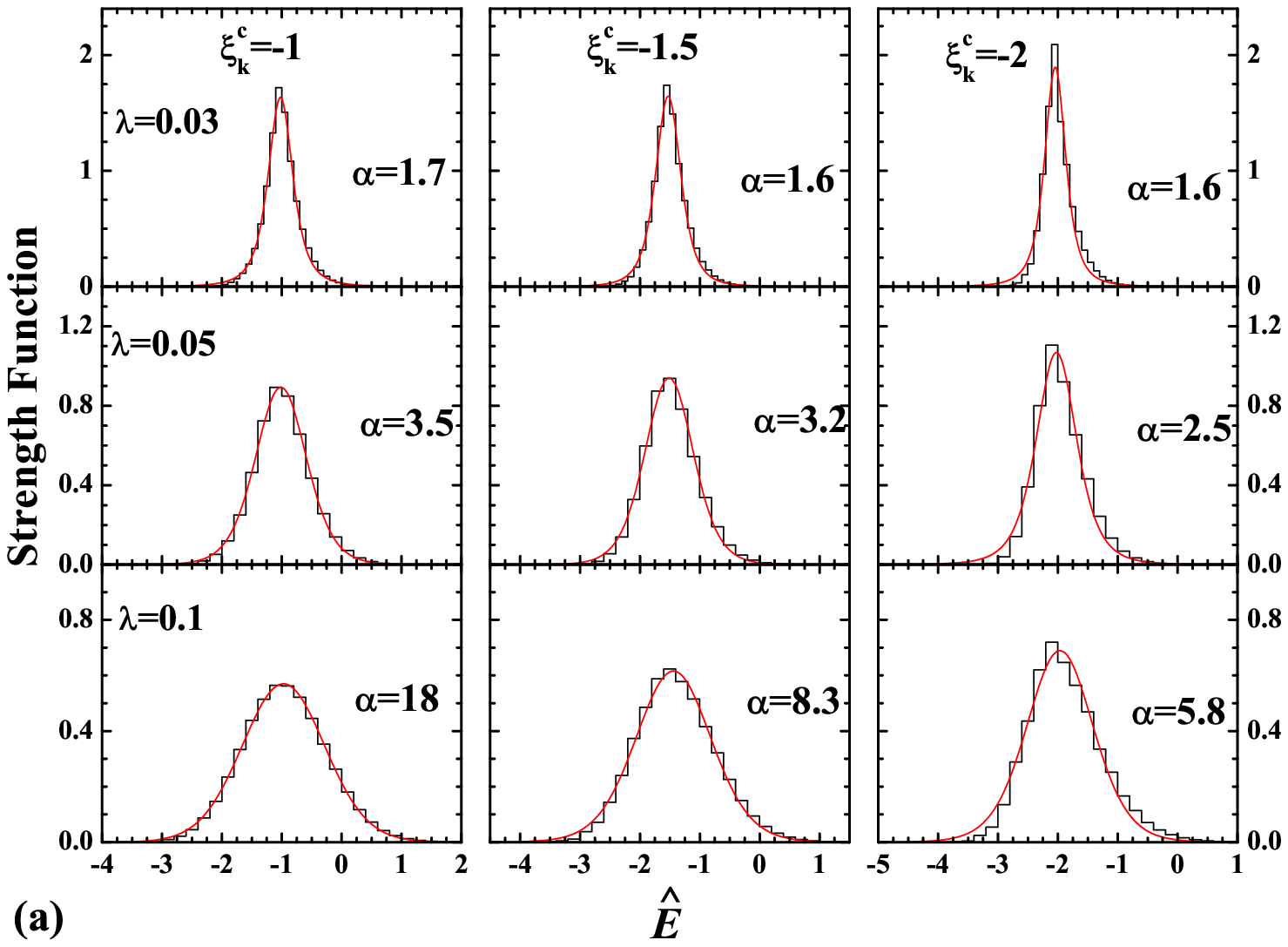}\\
\includegraphics[width=\linewidth]{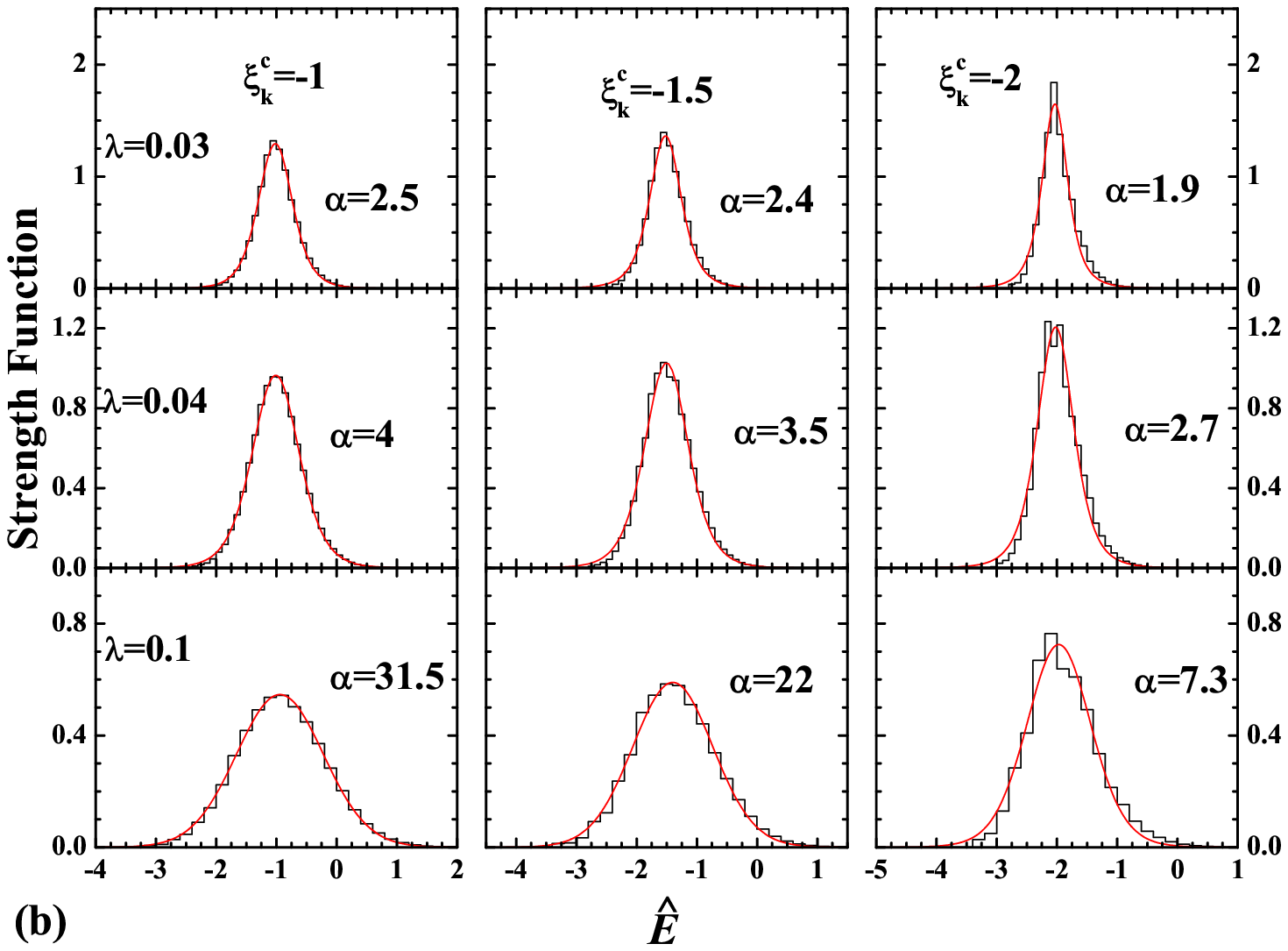}
\end{center}
\caption{Histograms represent strength functions vs. normalized energy, for three different values of
$\lambda$ using a 100 member BEGOE(1+2)-$F$ ensemble with $\Omega=4$, $m = 10$ system. Strength function plots are obtained for
$\hat{\xi^c_k} =-1, -1.5$ and $-2$ and results are shown for spins (a) $F=3$ and (b) $F=5$. Continuous curves in the figures correspond to
the $t$-distribution given by Eq.~(\ref{fkstu}). See text for further details. Results for $\hat{\xi^c_k} =1, 1.5$ and $2$ are not shown in the figure as they are similar to those for $\hat{\xi^c_k} =-1, -1.5$ and $-2$ respectively.}
\label{fke-12}
\end{figure}

The effects of localization-delocalization transition in the eigenfunctions
reflect in the degree of mixing of the eigenfunctions of the system.
Eigenfunction structure is understood usually in terms of the shape of the
strength functions or local density of states $F_k(E)$ where $k$ is an initial
basis state in which the system is prepared. Note that (see Appendix-A for
details) $F_k(E)$ gives spreading of the $k$ states over the eigenstates, i.e.
distribution of the square of the expansion coefficients $C_k^E$ in $\l.|k\ran =
\sum_E C_k^E \l.|E\ran$. In the BEGOE(1+2)-$F$ study,  we take the $k$ states to
be the eigenstates of both $h(1)$ and $F^2$ operators; see
\cite{KCS2006,Tureci2006,kaplan2000,Jacquod2001,papenbrock2002}. Our interest is
to obtain the ensemble averaged form of the strength functions. Towards this
end, eigenvalues $E$ and the $k$-energies $\xi_k$ ,for each member, are made zero
centered and scaled to the width [$\sigma_H(m,F)$] of the eigenvalue spectrum.
The new energies are called $\hat{E}$ and $\hat{\xi_k}$, respectively. Now, for
each member, all ${|C_{k}^{E}|}^2$ are summed over the basis states
$k$ with energy $\hat{\xi_k}$ in the energy window
$\hat{\xi^c_k} \pm \Delta_k$. Then, the ensemble averaged $F_k(\hat{E})$ vs
$\hat{E}$ are constructed as histograms by applying the normalization condition
$\int F_k(\hat{E})d\hat{E}=1$. We have chosen $\Delta_k=0.025$ for $\lambda  <
0.06$ and beyond this $\Delta_k=0.1$.  Figure~\ref{fke} shows
ensemble averaged strength functions $F_k(E)$ for the $k$ states averaged over
$\hat{\xi^c_k} - \Delta_k \leq \hat{\xi_k} \leq \hat{\xi^c_k} + \Delta_k$ with
$\hat{\xi^c_k} =0$. Results are shown for various values of $\lambda$ for the
system with $m=10$ bosons in $\Omega=4$  sp orbitals and spin $F=0$, 3 and 5.
Similarly, Figure~\ref{fke-12} shows ensemble averaged strength function for $\hat{\xi^c_k} =-1,- 1.5$ and
$- 2$ with three different values of $\lambda$ and for spins $F=3$  and $F=5$. The results clearly display transition from BW form to the Gaussian form, as the strength of the two-body interaction $\lambda$
increases beyond $\lambda_c(F)$. With $h(1)$ alone, that is for $\lambda = 0$, the basis
states become the eigenstates and therefore the strength
functions are $\delta$-functions. With increase in $\lambda$, the states begin
to spread and configurations begin to mix due to the two-body interaction, and
the strength functions acquire BW form (see Eq.~\ref{eq.app3}). With further
increase in $\lambda$, the strength functions start becoming
Gaussian in nature (see Eq.~(\ref{eq.app4}). Note that the BW form implies weak
de-localization and Gaussian form implies full delocalization. For BEGOE (or
EGOE), full delocalization (or fully chaotic) is in a energy shell but not in
the whole unperturbed basis and this aspect will be elaborated at the end of the
next Section.

In order to quantify the BW to Gaussian transition, used is the
BW to Gaussian interpolating student-$t$ distribution with a shape parameter
$\alpha$; see Eq.~(\ref{fkstu}) and Refs.~\cite{Angom2004,Ma-PRE} for more
details regarding the $t$-distribution. The continuous red curves in
Fig.~\ref{fke} are obtained by fitting the strength function histograms with
Eq.~(\ref{fkstu}) and the values of the shape parameter $\alpha$ are given in
the figure. Note that the results in Fig.~\ref{fke} are for $\hat{\xi^c_k}=0$. As seen from the results, the fits are excellent over a wide range
of $\lambda$ values. The criterion $\alpha \sim 4$ defines the transition point
$\blamf$  \cite{Angom2004}. Value of the parameter $\alpha$ increases slowly up
to $\blamf$, then it increases sharply. For $\alpha > 16$, the curves are
indistinguishable from Gaussian. From the results in Fig.~\ref{fke}, it is seen
that the transition point $\blamf$ is close to 0.06, 0.05 and 0.04 for $F$=0, 3
and 5 respectively. For a qualitative understanding of the variation of $\blamf$
with $F$ spin, we use the procedure adopted in the study of EGOE(1+2)-$\cs$
\cite{Ma-PRE}, where the spreading width of the strength functions and the
participation ratio (PR) are used to obtained the chaos marker in terms of the
variance propagator \cite{Ma-PRE}. For BEGOE(1+2)-$F$ we have,

\be
\blamf(F) \propto \dis\sqrt{\dis\frac{m\Omega^2}{Q(\Omega,m,F)}}\;\;.
\label{eq.lf}
\ee
Thus, the marker $\blamf$ is essentially determined by the variance propagator
$Q(\Omega,m,F)$. From the results in Fig.~6 in \cite{Ma-F}, it is clear that
$\blamf$ should decrease with $F$. This prediction is in agreement with the
results shown in Fig.~\ref{fke}. Therefore, Eq.~(\ref{eq.lf})  gives a good
qualitative understanding of $\blamf(F)$ variation with $F$ just as for
$\lambda_c(F)$ shown in \cite{Ma-F}. Turning to Figure~\ref{fke-12}, it can be argued
that,, for the BEGOE system considered, the BW to Gaussian transition extends upto $\hat{\xi^c_k}=\pm2$. From $-2$ and
below the transition is  slower and strength function is not symmetric.
Due to  the finiteness of the system, number of states below $\hat{\xi^c_k}=-2$
 will be much smaller than those above $-2$. Also, as seen from Fig.~\ref{fke-12},  $\blamf(F)$ depends on $\hat{\xi^c_k}$ and its value increases as we go away from the center.

In order to understand further the
structures due to delocalization of the eigenfunctions, we will consider
the two standard measures, the participation ratio and
information entropy in the next section; these are defined in Appendix-A.

\section{Participation ratio and Information entropy}
\label{sec4}

\begin{figure}[t]
\begin{center}
\includegraphics[width=\linewidth]{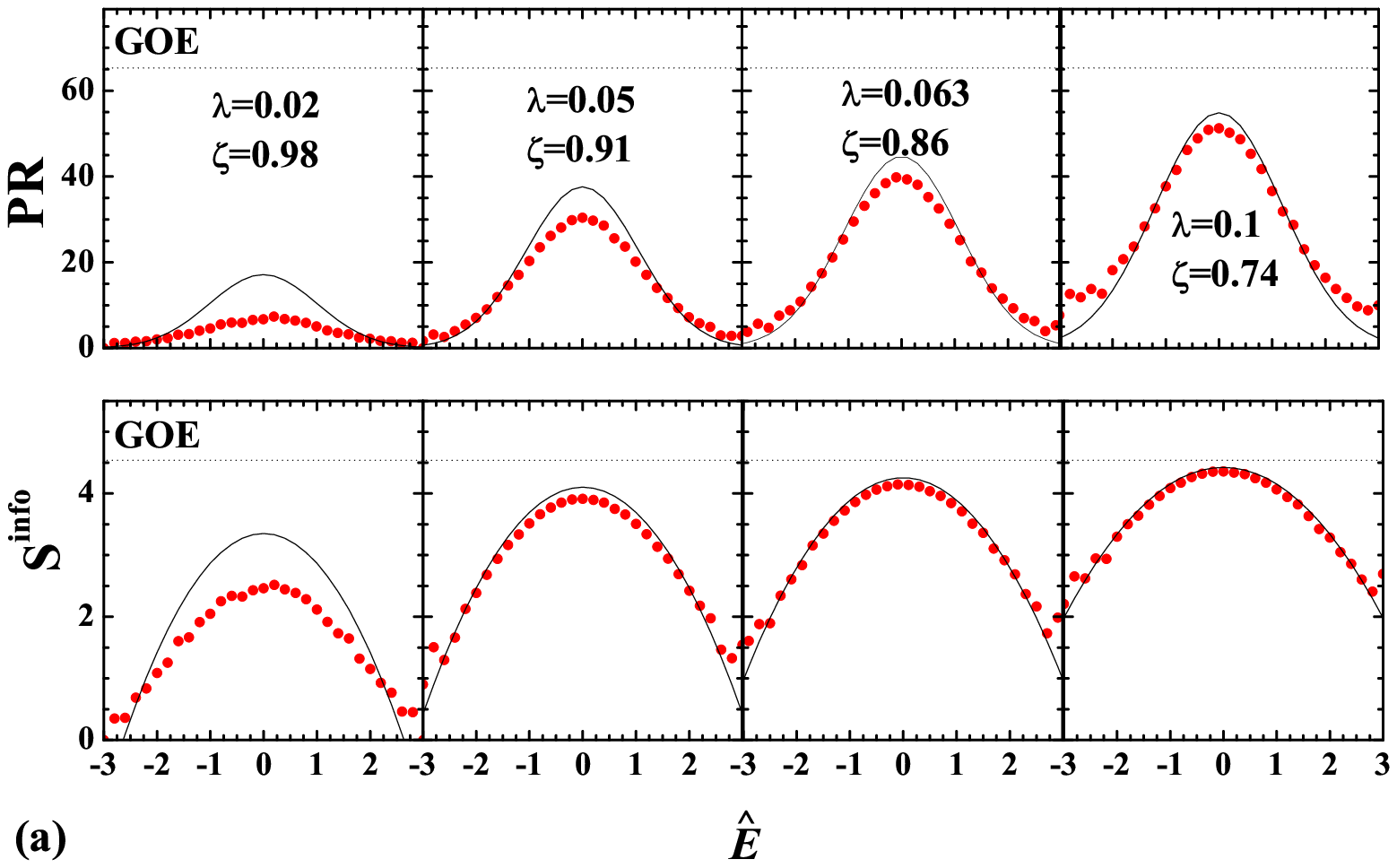}\\
\includegraphics[width=\linewidth]{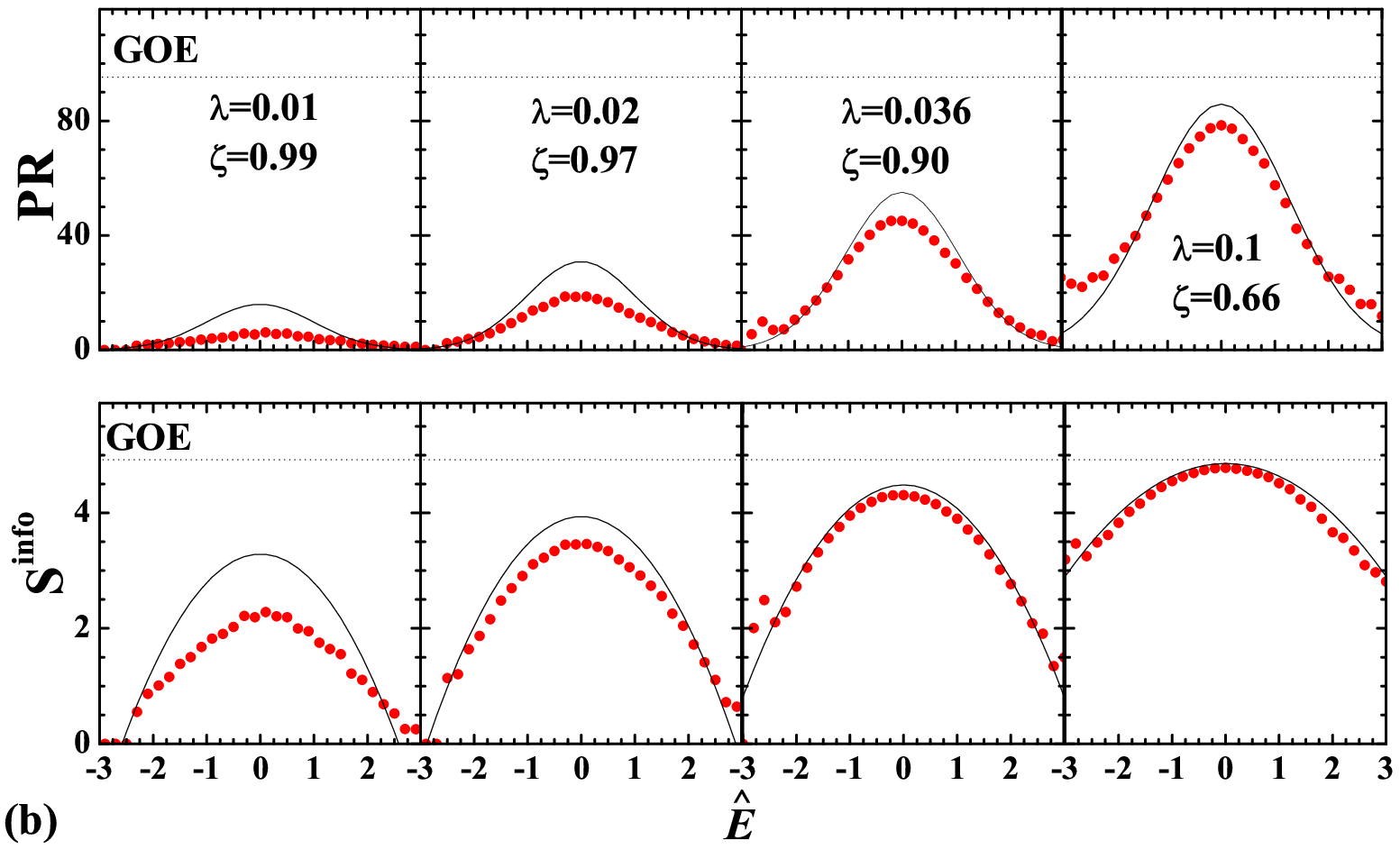}
\end{center}
\caption{ PR and $S^{\mbox{info}}$ vs. normalized energy $\hat{E}$ for a
system of $m=10$ interacting bosons in $\Omega=4$ sp orbitals. Results are shown
for (a) minimum spin $F=0$ and (b) maximum spin $F=5$. In each case 100 members
are used. Ensemble averaged results are represented by red circles while
continuous curves correspond to the theoretical estimates in the chaotic domain
as given by Eqs.~(\ref{eq.npc}) and (\ref{eq.sinfo}).}
\label{npcsinfo}
\end{figure}

Participation ratio (PR) or number of principal components is a useful quantity
to measure the degree of delocalization of a given eigenstate. The
$\mbox{PR}(E)$ defined by Eq.~(\ref{eq.npcc}) gives essentially the number of
basis states $k$ that make up the eigenstate with energy $E$.
Fig.~\ref{npcsinfo} shows the variation of PR for the same 10
boson system used in generating Figs.~\ref{fke} and \ref{fke-12} and for spins $F=0$ and 5. In
the figures, continuous curves correspond to the BEGOE(1+2)
formula given by Eq.~(\ref{eq.npc}) and the numerical ensemble averaged PR
values are shown as red filled circles. One can see that for small values of
$\lambda$, where the one-body part of the interaction is dominating with the
spacing distribution is close to Poisson in character, the theoretical curve is,
as expected, far away from the numerical results. With $\lambda =\lambda_c$
also, corresponding to the transition from Poisson to Wigner distribution for
NNSD, the theoretical estimate for PR is above the ensemble averaged values.
However, the ensemble averaged results are close to those given by
Eq.~(\ref{eq.npc}) for $\lambda$ near or above $\blamf$. Thus Eq.~(\ref{eq.npc})
is good in the de-localized regime with $\lambda \gazz \blamf$.

Another statistical quantity closely related to PR is the information entropy
$S^{\mbox{info}}(E)$ in the eigenfunctions; Eq.~(\ref{eq.app6}) gives the
definition. Clearly, increase in information entropy implies
more delocalization of the eigenstates. It is well demonstrated in
\cite{casati98,horoi95} that the  thermodynamics entropy defined by the state
density, the information entropy in the eigenfunctions expanded in the
mean-field basis and the single particle entropy defined by the mean occupation
numbers of the sp states, all coincide for strong enough interaction but only in
the presence of a mean field; i.e. in the chaotic domain ($\lambda > \blamf$)
but with a mean field. In Fig.~\ref{npcsinfo}, numerical results for
$S^{\mbox{info}}$ for the 10 boson system with $F=0$ and $5$ are compared with
the results from the BEGOE(1+2) formula (see Eq.~(\ref{eq.sinfo})) for the same
values of $\lambda$ as used for PR. As seen from the figures,
there is a good correlation between $S^{\mbox{info}}$ and PR results as expected
\cite{Angom2004,Jacq-02,Varga}. Also, for $\lambda \gazz \blamf$,  theoretical
results given by Eq.~(\ref{eq.sinfo}) are close to the ensemble averaged
numerical results for different $F$ values. This again confirms
that as $\lambda$ increases beyond $\blamf$ eigenstates start becoming fully delocalized (chaotic)
 and the meaning of full delocalization is discussed below. From
the PR and $S^{\mbox{info}}$ results in Fig.~\ref{npcsinfo}, it
is seen that the transition point $\blamf$ deduced in Section \ref{sec3} corresponds to
$\zeta \sim 0.75-0.8$.

Results in Fig.~\ref{npcsinfo} show that even when $\lambda >>
\blamf$, the eigenstates will not occupy the whole unperturbed basis and this
is due to the fact that the interaction is two-body in nature (i.e. $2 << m$).
In this situation, one can use the notion of delocalization in a 'enery shell'
whose width is defined by the interaction. Because of this, PR and information
entropy in general will not reach the value given by full random matrices. This
is confirmed by the results in Fig.~\ref{npcsinfo} and also by Eqs.~(\ref{eq.npc},\ref{eq.sinfo}).
It should noted that even for energies taken not close to the center,  the
eigenstates can be fully chaotic and delocalized in the energy shell, and not in
the whole unperturbed basis. Quantitative results for the notion of energy shell
and the width of the energy shell are obtained using banded random matrix
ensembles by Casati et al. \cite{Cas-1,Cas-2}; see also the discussion in
\cite{BISZ2016}. For EGOE(1+2) or BEGOE(1+2), given a eigenstate with energy
$E$, an estimate of the location and width of the energy shell follows from the
results given in \cite{vkbrs01}. We will discuss this in the following section
after examining the structure of the eigenstates with $\lambda$ increasing
beyond $\blamf$.

\section{Thermodynamic region: $\lambda_t$ marker}
\label{sec5}

\begin{figure}[t]
\begin{center}
\includegraphics[width=0.6\linewidth]{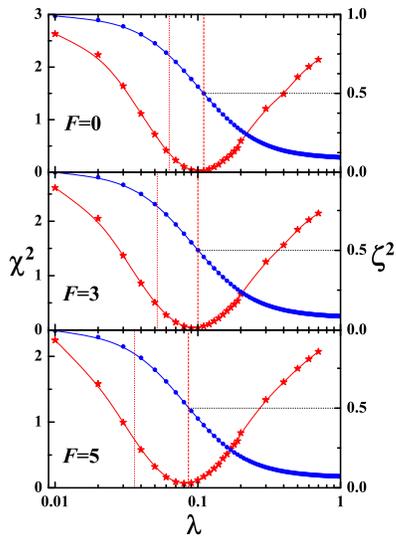}
\end{center}
\caption{ Variation of $\chi^2$ (red stars) and $\zeta^2$ (blue circles)  as
a function of the interaction strength $\lambda$, calculated for a 100 member
BEGOE(1+2)-$F$ ensemble with  $\Omega=4$,$m = 10$ are shown for $F$ spin values
0, 3 and 5. The $\lambda_t$ marker for the thermodynamic region corresponds to
$\chi^2=0$ and $\zeta^2 = 0.5$. The vertical red doted and dashed lines in each panel indicate the respective
positions of $\blamf$ and $\lambda_t$.}
\label{duality}
\end{figure}

The thermodynamic region defined by $\lambda \sim \lambda_t$ is the region where
any definition of the thermodynamic  quantities such as
entropy, temperature etc. gives the same result. Then, one can argue that in
this region all wave functions look alike (similar to ETH). In fact, this is
also the duality region for embedded ensembles \cite{Jacq-02,Angom2004,Ma-PRE}.
Just like for other embedded ensembles, for the BEGOE(1+2)-$F$ Hamiltonian,
there are two choices of basis appear naturally. One is the mean-field basis
defined by $h(1)$ and another is the infinite interaction strength basis defined
by $V(2)$. To locate duality or equivalently for establishing the thermodynamic
region, we compute information entropy $S^{\mbox{info}}(E)$ in these two basis.
At $\lambda=\lambda_t$, the spreadings produced by $h(1)$ and $V(2)$ are equal
giving the correlation coefficient $\zeta$ to be $\zeta^2(\lambda_t)=1/2$
\cite{Angom2004}; see Eq. (\ref{eq.app5}) for the definition of $\zeta^2$ and
Eqs.~(\ref{eq.npc}) and (\ref{eq.sinfo}) for the role of $\zeta^2$ in PR and
$S^{\mbox{info}}(E)$. Hence, putting $\zeta^2=1/2$ in Eq.~(\ref{eq.sinfo}) gives
the basis independent expression $S_{\lambda_t}^{\mbox{info}}(E)= \ln
(0.48d)+\ln \sqrt{1/2} +\frac{(1-E^2)}{4}$. To identify $\lambda_t$, we use a
$\chi^2$ measure giving sum of the squares of the deviations between numerically
obtained $S^{\mbox{info}}(E)$ in $h(1)$ basis and $V(2)$ basis with the
expression (see Eq.~\ref{eq.sinfo})for $S_{\lambda_t}^{\mbox{info}}(E)$:

\be
\chi^2=\dis \int_{-\infty}^{\infty} \mathrm{d}E\;\l[ { \{R_{\lambda_t}^E-R_{h(1)}^E\}^2 +
\{R_{\lambda_t}^E-R_{V(2)}^E\}^2 }\r]\;.
\ee
Here $R_{\alpha}^E=\exp[S_{\alpha}^{\mbox{info}}(E)- S^{\mbox{info}}_{GOE}]$
with $\alpha$ representing $h(1)$ or $V(2)$. The minimum value of $\chi^2$ gives
the $\lambda_t$. Figure~\ref{duality} shows variation of $\chi^2$ (red stars)
with the interaction strength $\lambda$ for a 100 member BEGOE(1+2)-$F$
ensemble. Used here is the ensemble for  $\Omega=4$,$m = 10$ and the results are
shown for $F$ spin values 0, 3 and 5. It can be seen that for the present
example, $\lambda_t \simeq 0.11$ for $F$=0, $\lambda_t \simeq 0.10$ for $F$=3
and $\lambda_t \simeq 0.086$ for $F$=5. The two vertical doted lines, in each
plot, indicate the respective positions of $\blamf$  and $\lambda_t$. For
$\lambda < \lambda_t$, the $S_{h(1)}^{\mbox{info}}$ values are higher and
$S_{V(2)}^{\mbox{info}}$ values are smaller compared to
$S_{\lambda_t}^{\mbox{info}}$  and for  $\lambda > \lambda_t$, the situation is
reversed giving $\chi^2 > 0$.  The values of entropy, in the two basis coincide
with $S_{\lambda_t}^{\mbox{info}}$ at $\lambda=\lambda_t$ giving $\chi^2\sim0$.
In Fig.~\ref{duality}, ensemble averaged values of $\zeta^2$ (blue circles) are
also shown for each $F$ spin. It is clear from the figure that for smaller
$\lambda$ ($\lambda \leq \lambda_c$), $\zeta^2$ is close to $1$ and as $\lambda$
increases, $\zeta^2$ goes on decreasing smoothly. It is seen
from the figure that the condition $\zeta^2(\lambda_t)=0.5$ gives the same
values for the marker $\lambda_t$ as obtained using information entropy.

Qualitative understanding of the variation of $\lambda_t$ with $F$ is obtained
using the fact that at $\lambda=\lambda_t$ we have
$\zeta^2(\lambda_t)=0.5$. This leads to the condition $\sigma^2_{h(1)}(m,F) =
\lambda^2_t\;Q(\Omega,m,F)$. Then, for BEGOE(1+2)-$F$, using
Eq.~(\ref{eq.sig2h1}) for $\sigma^2_{h(1)}$ will give,

\be
\lambda_t(F) \propto \dis\sqrt{\dis\frac{\sigma^2_{h(1)}(m,F)}
{Q(\Omega,m,F)}}\;.
\label{eq.inf4}
\ee
Eq.~(\ref{eq.inf4}) gives a good qualitative understanding of $\lambda_t$
variation with $F$. Before going further, let's add that even
when $\lambda$ value is large, numerical results in Fig.~\ref{npcsinfo} show that close to the bottom
of  the spectrum thermalization is absent and the corresponding eigenstates
are not fully chaotic. For a more  complete  understanding of the structure of the ground  state  region  generated by  embedded ensembles, numerical results with  much larger values of $(m,\Omega)$ are needed and this is future studies.

Returning to the energy shell mentioned at the end of Section~\ref{sec4}, in the
thermodynamic region, form of the strength functions will be Gaussian and the
eigenstates are fully (chaotic) delocalized in the energy shell. Using Eq.~(1)
of \cite{vkbrs01}, we have the  result that the distribution of $\l|C^E_k\r|^2$ vs $\xi_k$ for
a fixed $E$ is a Gaussian, in the thermalization region, with center at $\zeta\hat{E}$ and width $\sqrt{1-\zeta^2}$. Note
that $1-\zeta^2$ is the spectral variance generated by the two-body part of the
Hamiltonian (measured in units of the total spectral variance). As $\zeta^2=1/2$ in the thermalization region,  it can be argued that the width (energy span) of the energy shell is  $\sim 2\sqrt{1-\zeta^2}\sigma_H=\sqrt{2} \sigma_H$. Moreover, for $\lambda \leq \blamf$ as discussed before $\zeta^2 >> 1/2$ and therefore the eigenstates are much less delocalized than in the thermodynamic
region. For a better quantitative understanding of the energy shell in embedded
ensembles calls for a more rigorous mathematical treatment of these ensembles
and/or numerical investigations with much larger values for $(m,\Omega)$. Clearly, this needs to be investigated but it
 is outside the scope of the present paper.

In summary, the above results along with the previous studies, for EGOE(1+2),
EGOE(1+2)-$\cs$ and BEGOE(1+2), establish universality of the BW--Gaussian transition
in strength functions and the  region of thermalization marked by $\blamf$  and
$\lambda_t$ respectively, both for bosonic and fermionic systems (with and
without good quantum numbers). Similar structures were obtained using spin
models employed by Lea Santos and others \cite{Rigol16,BISZ2016} and thus it is
plausible to conclude that embedded ensembles are generic models for isolated
finite interacting many-particle quantum systems.

Going further, in the next section, we will study entanglement entropy within
embedded ensembles for spin-less boson systems. The entanglement measures,
introduced in the context of Quantum Information Science, are used to
characterize complexity in quantum many-body systems. Entanglement and
delocalization are found to be strongly correlated for disordered spin-1/2
lattice systems \cite{Br-08,Pi-08}.

\section{Entanglement entropy in embedded ensembles}
\label{sec6}

The bipartite entanglement entropy ($S^{EE}$) is a measure of the strength of
quantum correlations between two parts of a many-body system. Consider an
eigenstate with energy $E$, $| E \rangle$. Then, the full density matrix is given by $\rho(E) =
| E \rangle \langle E |$. Let the system is divided into two parts $A$
and $B$ giving partition of its Hilbert space $\bhh = \bhh_A \otimes \bhh_B$,
where  $\bhh_A$ and $\bhh_B$ are the Hilbert subspaces respectively. Now, an
eigenstate is said to be separable if it can be written as $|E\rangle =
|a_E\rangle  \otimes |b_E\rangle$. Here, $|a_E\rangle$ and $|b_E\rangle$ are the
basis states residing in the Hilbert subspaces $\bhh_A$ and $\bhh_B$
respectively. A state $|E\rangle$ is said to be entangled if it is not
separable. The $S^{EE}$ between the partitions $A$ and $B$, in the eigenstate
$|E\rangle$, is given by,

\be
S^{EE}(E) = -Tr \rho_A(E) \log {\rho_A(E)} = -\sum_\mu \tau_{\mu}(E) \log
{\tau_{\mu}(E)}.
\ee
Where $\rho_A(E) = Tr_B \rho(E)$ is the reduced density matrix of the $A$ part,
obtained from the full density matrix $\rho(E)$ by tracing out $B$ degrees of
freedom and $\tau_{\mu}(E)$ are the eigenvalues $\rho_A(E)$. For a separable
state $S^{EE}=0$ while for a maximally entangled state $S^{EE}_{max}=\log
d_{min}$, where $d_{min}= \min(d_A,d_B)$; $d_A$ and $d_B$ are the
dimensionalities of the Hilbert subspaces $\bhh_A$ and $\bhh_B$ respectively. As
in the definition of the $\rho(E)$, both self-correlations as well as
cross-correlations between the coefficients $C^E_{k}$ are taken into account,
unlike the PR and information entropy where the cross-correlations are
neglected, the $S^{EE}$ is independent of the  basis states. However, it does
depend on the partition.

\begin{figure*}[!th]
\begin{center}
\includegraphics[width=0.55\textwidth]{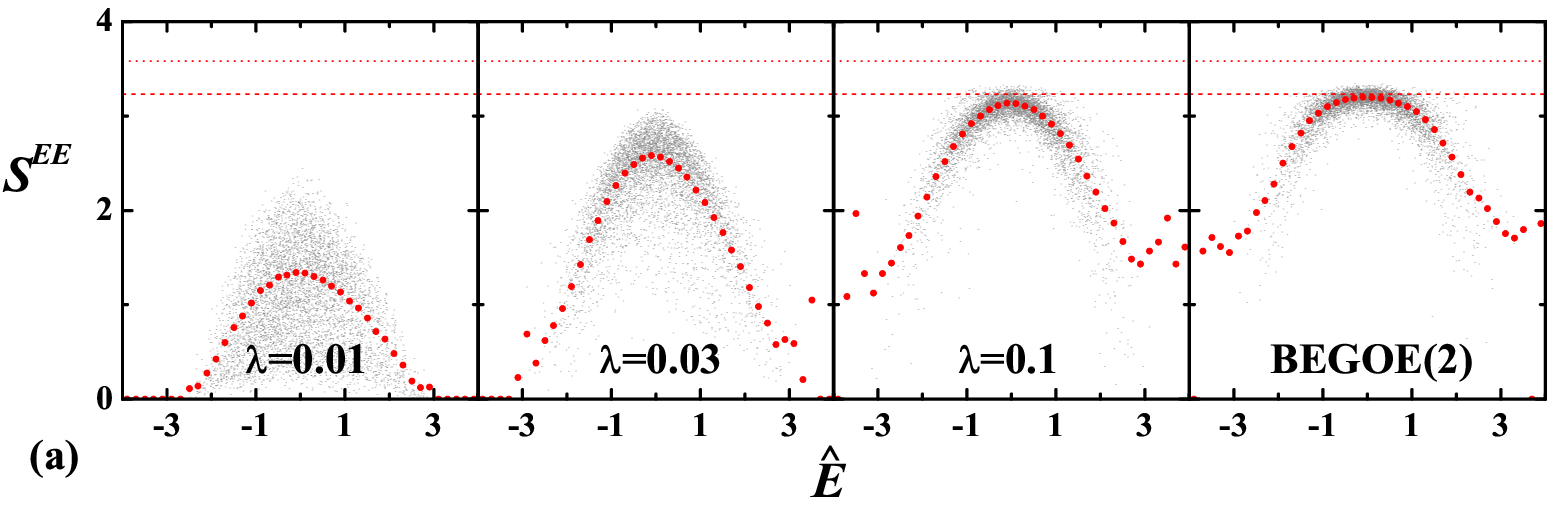}\\
\includegraphics[width=0.55\textwidth]{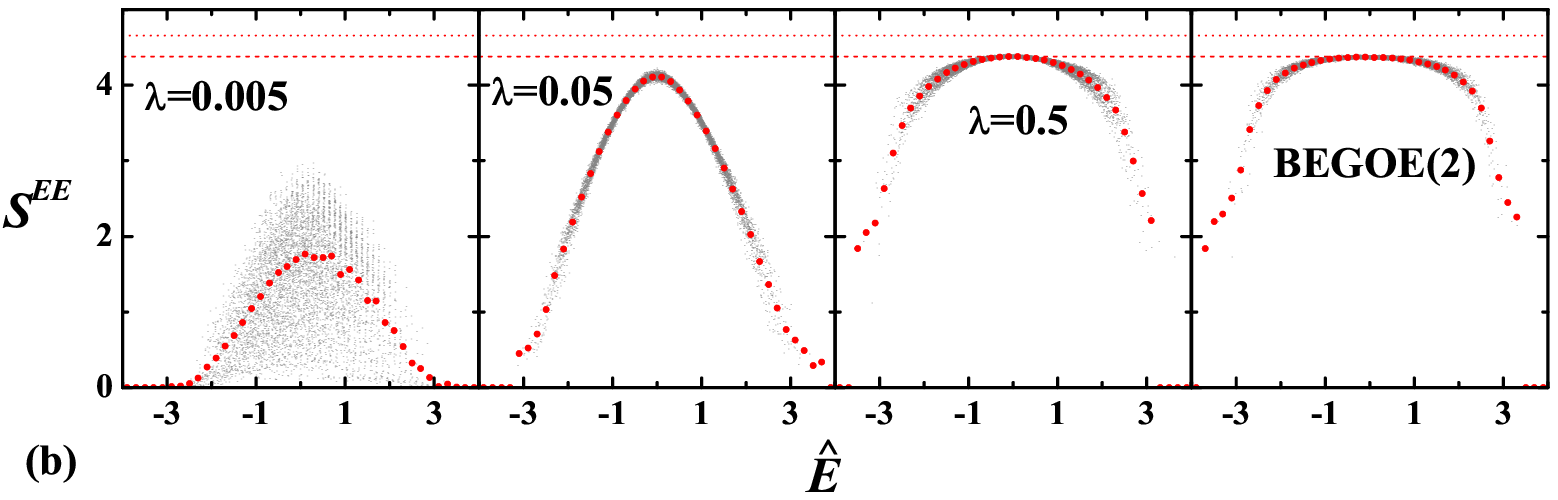}
\end{center}

\caption{\label{pree} Ensemble averaged entanglement entropy ($S^{EE})$ vs.
normalized energy $\hat{E}$ for various values of $\lambda$ are shown by red
circles; (a) with $m=10$ spin-less bosons in $N=4$ sp states and (b) with $m=8$
spin-less bosons in $N=8$ sp states. In (a) each case 100 members are used and
in (b) each case 5 members are used. The ensemble average is carried out by
making the spectra of each member of the ensemble zero centered and scaled to
unit width. Data points of a few members are also shown (small dots). The
entanglement partition is between two equal halves of the sp states. The red
dashed and red doted lines indicate the $S^{EE}_{GOE}$ and maximal $S^{EE}$,
$S^{EE}_{max}$ respectively. For ($m=10,N=4$) example, $S^{EE}_{GOE}=3.23$ and
$S^{EE}_{max} = \log {36}$, while for ($m=8,N=8$) example, $S^{EE}_{GOE} = 4.38$
and $S^{EE}_{max} = \log{105}$. The last plots in both (a) and (b) are for
BEGOE(2) only.}

\end{figure*}

In Figure~\ref{pree} shown are ensemble averaged results for $S^{EE}$ as a
function of the normalized energies for various values of $\lambda$ in $H=h(1) +
\lambda V(2)$. The systems considered are: (a) $m=10$ spin-less bosons in $N=4$
sp states; (b)  $m=8$ spin-less bosons in $N=8$ sp states. See \cite{Ch-0304}
for details of these BEGOE(1+2)ensembles. The last panel in both
Fig.~\ref{pree}(a) and \ref{pree}(b) are for BEGOE(2) only. In the present study, the $S^{EE}$ is
obtained between two equal partitions of sp states. Then, $C^E_k$ changes to $C^E_{k_A , k_B}$ and
$\rho_A(E)$ is given by $\sum_{k_B}\,C^E_{k_A , k_B} C^E_{k^\prime_A k_B}$.  It
can be clearly seen from Fig.~\ref{pree} that for small $\lambda$, small
$S^{EE}$ values are found in the tails of the spectrum while large as well as
small $S^{EE}$ values are found in the middle part of the spectrum. As $\lambda$
increases, the $S^{EE}$ varies smoothly with energy. For sufficiently large
$\lambda$, large $S^{EE}$'s in the middle part and small $S^{EE}$'s in the tails
of the spectrum are found. This behavior can be understood as follows: For
$\lambda=0$, with $h(1)$ part only, the basis states ($| k \rangle$)
themselves are the eigenstates ($| E \rangle$) and hence the eigenstates are
separable giving $S^{EE}=0$ for all the eigenstates. As soon as the interaction
is switched on, the basis states begin to spread and the configurations start to
mix due to the two-body interaction. At this point, the cross-correlations in
the $\rho(E)$ start growing leading to mixing of the subspaces weakly and hence
resulted into entangled eigenstates with non-zero but low $S^{EE}$ values. As in
this region, the structure of eigenstates is not chaotic enough, leading to
strong variation in $C^E_{k}$ and thus in the $\rho(E)$ leading to stronger
variation in $S^{EE}$ values over the ensemble and particularly in the middle
part of the spectrum.  Then, one may argue that the eigenstates near the tails
show area-law behavior, while those in the middle part show both area-law and
volume-law behavior; for area and volume-laws for $S^{EE}$ see \cite{ee-area}.
With further increase in $\lambda$, more and more basis states contribute to the
eigenstates leading to increase in $S^{EE}$ values. Here, reduction in the
fluctuations leads to smooth variation of $S^{EE}$ with energies. For
sufficiently stronger $\lambda$, the eigenstates, in the middle of the spectrum,
become fully chaotic and the cross-correlations in the $\rho(E)$ are stronger
enough to mix the subspaces completely giving maximum $S^{EE}$ value. As chaos
sets in fast in the eigenstates in the middle part of the spectrum compared to
the region near the ground state, larger values of $S^{EE}$ are generated in the
middle part of the spectrum and small $S^{EE}$ values in the tail region of the
spectrum. This suggests that the eigenstates have area-law like character near
the spectral tails while volume-law like character in the middle part.  The
results here are in good agreement with those obtained in
\cite{alba09,wouter2015}.

\begin{figure*}[!th]
\begin{center}
\includegraphics[width=0.55\textwidth]{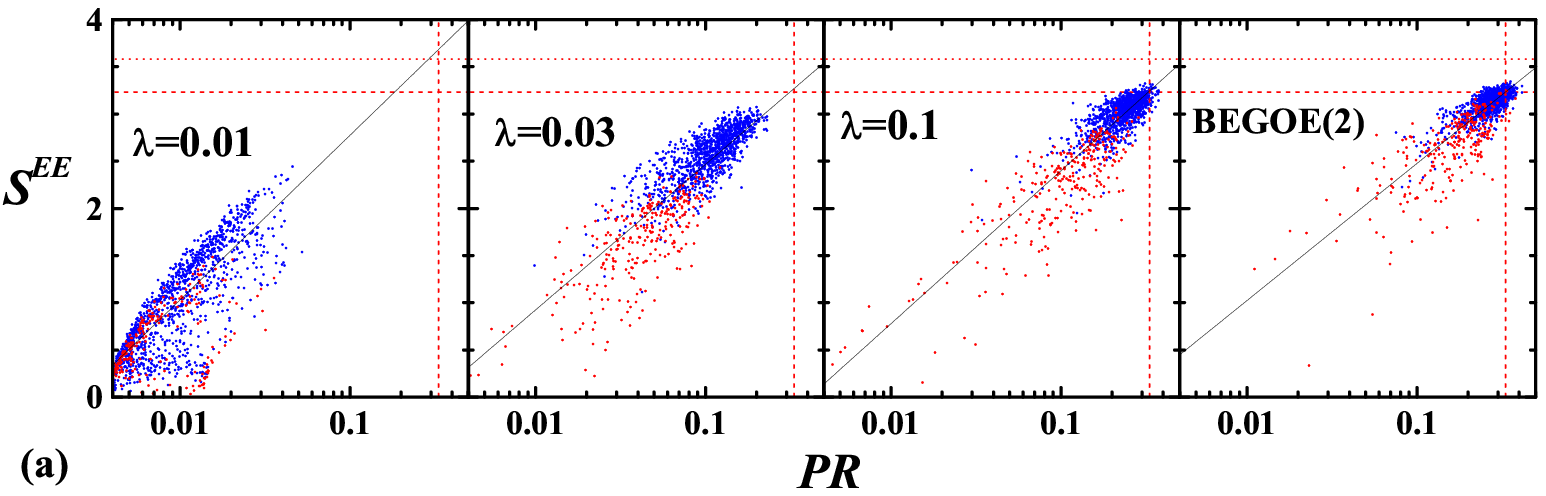}\\
\includegraphics[width=0.55\textwidth]{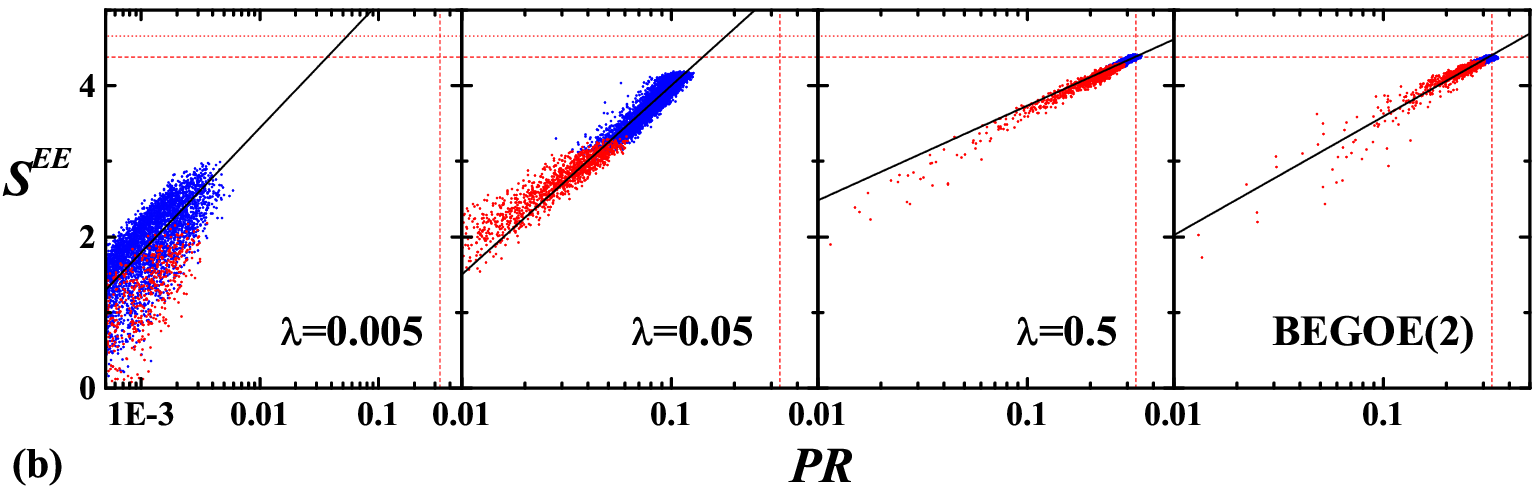}
\end{center}
\caption{ Scatter plots of entanglement entropy
$S^{EE}$ vs. participation ratio (PR) for (a) with $m=10$ spin-less bosons in $N=4$ sp states and (b) with
$m=8$ spin-less bosons in $N=8$ sp states. Data points for a single member of
the ensembles are shown. The blue points indicate middle 80\% eigenstates of the
spectrum while the red dots represent 10\% eigenstates from either side of the
tails of the spectrum. The vertical and horizontal red dashed lines indicate $\mbox{PR}_{GOE}$
and $S^{EE}_{GOE}$ respectively while the horizontal red doted lines indicate $S^{EE}_{max}$. The black straight lines are due to ODR fits to
the full data sets.}
\label{fig6}
\end{figure*}

Following \cite{wouter2015}, we are further motivated to compare PR and
$S^{EE}$. In Fig.~\ref{fig6}, the scatter plots of $S^{EE}$  vs. PR for all the
eigenstates are shown for one typical member of each of the two ensembles
considered in Fig.~\ref{pree}. Note that $\mbox{PR}(E)=\{ d \sum_k
|C^E_k|^4\}^{-1}$ where $d$ is the dimension of the $H$ matrix. For
($m=4,N=10$), $d=286$ while for ($m=8,N=8$), $d=6435$. The PR values are plotted
on a logarithmic scale. In the plots, the middle 80\% eigenstates of the
spectrum are shown using blue dots while remaining 10\% of eigenstates from both
ends of the spectrum are shown using red dots. The difference between the nature
of middle part and the tails of the spectrum is clearly visible. For weak
interaction strength $\lambda$, both PR and $S^{EE}$ values are much smaller
than the random-state predictions for all the eigenstates. With increase in
$\lambda$, PR and $S^{EE}$ values drift towards the random-state expectation
values for more and more eigenstates from the middle part of the spectrum. With
further increase in $\lambda$, the eigenstates in the middle part become fully
chaotic leading to clustering of both PR and $S^{EE}$ near the random-state
expectation values. The eigenstates near both the tails of the spectrum have
much smaller values for these quantities. The straight lines in each of the the
Fig.~\ref{fig6} are obtained by Orthogonal Distance Regression (ODR) to the full
data sets \cite{wouter2015}. The BEGOE(1+2) results obtained in
Figs.~\ref{pree}(b) and \ref{fig6}(b) show strikingly similar structure as
obtained using Bose-Hubbard model in \cite{wouter2015}. This further confirms
that BEGOE generates generic results for localization-delocalization transitions
in finite interacting quantum systems.

\section{Conclusions:}
\label{sec7}

In the present paper, we have studied localization to delocalization transitions
using various traditional measures, based on the structure of eigenfunctions,
such as the shape of strength functions, participation ratio and information
entropy, as a function of the two-body interaction strength ($\lambda$) in
BEGOE(1+2)-$F$ Hamiltonian. It is demonstrated that BEGOE(1+2)-$F$ ensembles, in
addition to Poisson to GOE transition in level fluctuations at $\lambda_c$
\cite{Ma-F}, generate two more transition (chaos) markers namely, $\blamf$ ($>
\lambda_c$) and $\lambda_t$ ($ > \blamf > \lambda_c$) for BW to Gaussian
transition in strength functions and the duality or thermodynamic region
respectively. Also, the variance propagator $Q(\Omega,m,F)$ [given by
Eqs.~(\ref{eq.q1}) and (\ref{eq.q2})] gives a good qualitative explanation for
the $F$-spin dependence of these chaos markers. As $Q$ increases with $F$, and
using this in  Eqs.~(\ref{eq.lf}) and (\ref{eq.inf4}), establishes that $\blamf$
and $\lambda_t$ values will increase with $F$ just as $\lambda_c$ ($F=m/2$
corresponds to spin-less bosons). Hence, these results establish that the
introduction of the spin quantum number preserves the general structures,
generated by spin-less EGOE(1+2) (bosonic and fermionic) ensembles and
EGOE(1+2)-$\cs$ ensemble for fermions with spin, as established previously
\cite{Ma-PRE,Ko-14}, although the actual values of the markers vary with the
$m$-particle spin. In the literature, most of the results for localization to
delocalization transitions were obtained using spin models (with fermions and
hardcore bosons) and Bose-Hubbard model \cite{Rigol16,BISZ2016}. Now, we have
demonstrated that embedded ensembles give similar results and they do form a
good model to study localization to delocalization transitions.

In addition to the traditional measures, we have analyzed entanglement entropy
for spin-less BEGOE(1+2) ensembles using all the eigenstates and presented the
first results.  For weak interaction strength $\lambda$, the states with
low-entanglement and low-PR appear in whole part  of the spectrum. While for
sufficiently large interaction strength $\lambda$, larger values in the middle
part of the spectrum and smaller near the tails of the spectrum for both PR and
$S^{EE}$ are found.  Also correlations between PR and $S^{EE}$ are analyzed
qualitatively. All these  results are consistent with those obtained in
\cite{wouter2015} where Bose-Hubbard model has been employed. These show that
generic features can be well described by two-body ensembles (embedded
ensembles) as emphasized in \cite{Br-08}. One can also study ETH and many
aspects of thermalization using embedded ensembles. Recently, some of these are
numerically investigated using embedded ensembles in \cite{KoReta,Haldar2016}.
In addition, using the simpler EGOE(1) generated by a random one-body
Hamiltonian (with both diagonal and off-diagonal matrix elements) for fermions,
ETH has been proved analytically by computing correlations and entanglement
entropies \cite{Mag-16}. Hence, it is interesting and important to investigate
these using two-body and one plus two-body embedded ensembles. This exercise
will be carried out in future. It  is also possible to consider Hamiltonians
giving in some extreme limits regular features and intermediately chaotic
behavior by adding some other regular Hamiltonians to the EGOE Hamiltonian
considered in this paper. This will be addressed in future.

\begin{acknowledgements}
 Authors thank B. Chakrabarti and Manan Vyas for many useful discussions.
 NDC acknowledges support from Department of Science and Technology,
 Government of India [Project No: EMR/2016/001327].
\end{acknowledgements}

\renewcommand{\theequation}{A-\arabic{equation}}
\setcounter{equation}{0}
\section*{APPENDIX A}

Given a $m$-particle state  $\l| k \ran$ and its expansion in terms of the
eigenstates $\l| E \ran$ of the  Hamiltonian $H$ with eigenvalues $E$,
\be
\l| k \ran = \dis\sum_E C_{k}^{E} \l| E \ran\;.
\label{eq.app1}
\ee

The strength function that corresponds to the state $\l| k \ran$ is
\be
F_{k}(E) =\sum_{E'} {|C_{k}^{E'}|}^2 \;\delta(E - E') =
d\,\overline{|C_{k}^{E}|^2}\,\varrho(E)\;.
\label{eq.app2}
\ee
Here, $\overline{|C_{k}^{E}|^2}$ is the average of $|C_{k}^{E}|^2$ over the
degenerate $E$ states, $d$ is the dimension of the $m$-particle space and
$\varrho(E)$ is the normalized eigenvalue density.  Energy $\xi_k$ of the states
$\l| k \ran$ are given by $\xi_k=\langle k | H | k \rangle$, the diagonal matrix
element of $H$ in the $k$ basis. The BW form for the strength functions, with
$\Gamma$ denoting the spreading width, is
\be
F_k(E) = \dis\frac{\Gamma}{2\pi}\,\dis\frac{1}{(E-\xi_k)^2 +
\dis\frac{\Gamma^2}{4}}\;.
\label{eq.app3}
\ee
Similarly, with $\sigma^2$ giving the spectral variance, the Gaussian form is
\be
F_k(E)=\dis\frac{1}{\sqrt{2\pi}\;\sigma}\,\exp\, -\dis\frac{(E-\xi_k)^2}{
2\,\sigma^2} \;.
\label{eq.app4}
\ee
Transition in $F_k(E)$ from BW to Gaussian form can be described using
the student-$t$ distribution \cite{Angom2004,Ma-PRE} given by
\be
F_k^{stud}(\hat{E}) d\hat{E}=\frac{(\alpha \beta)^{\alpha-1/2} \Gamma(\alpha)}{
\sqrt{\pi} \Gamma(\alpha-1/2)} \frac{\mathrm{d}\hat{E}}{[(\hat{E}-\hat{\xi_k})^2
+\alpha \beta ]^{\alpha}}\;.
\label{fkstu}
\ee
Here $\alpha$ is the shape parameter with $\alpha =1$ giving BW and $\alpha
\rightarrow \infty$ giving Gaussian form. The parameter $\beta$ defines the
energy spread of the strength function.

For an eigenstate $|E \rangle$ spread over the basis states $|k\rangle$,
PR is defined by,
\begin{equation}
\mbox{PR}(E) = \left\{ {\sum\limits_k {\left| {C_{k}^{E} }
\right|^4 } } \right\}^{ - 1}\;.
\label{eq.npcc}
\end{equation}
The GOE value for PR is $d/3$, where $d$ is the dimensionality of the $H$
matrix. For EGOE(1+2), expression for PR in the region where strength functions
are close to Gaussian form is given by \cite{vkbrs01};

\be
\mbox{PR}\left(\hat{E}\right)=\frac {d}{3}\sqrt{1-\zeta^4} \
\exp\left\{\frac{-\zeta^2 \hat{E}^2}{1+\zeta^2}\right\}\;.
\label{eq.npc}
\ee
Eq.~(\ref{eq.npc}) applies to BEGOE(1+2)-$F$ and here the normalized energy
$\hat{E}=(E-E_c(m,F))/\sigma_H(m,F)$ where $E_c(m,F)$ is the energy centroid in
$(m,F)$ space and $\sigma_H(m,F)$ is the spectral width generated by $H$.
Similarly, $\zeta$ is the correlation coefficient between the full Hamiltonian
$H$ and the diagonal part of $H$. With $H$ defined by Eq.~(\ref{eq-1}), we have
\cite{vkbrs01};

\be
\zeta = \sqrt{\frac{\sigma^2_{h(1)}(m,F)}{\sigma^2_{h(1)}(m,F)+\lambda^2
\sigma^2_{V(2)}(m,F)}}\;\;.
\label{eq.app5}
\ee
Clearly, as $\lambda$ the strength of the two-body interaction increases,
$\zeta$ will go on decreasing.  Similarly, EGOE(1+2) formula for the
information entropy ($S^{\mbox{info}}$) defined by,
\be
S^{\mbox{info}} (E)= - \sum\limits_k {\left| {C_{k}^{E} } \right|^2 }
\ln \left| {C_{k}^{E}} \right|^2 \;,
\label{eq.app6}
\ee
is given by \cite{vkbrs01},
\be
\barr{l}
S^{\mbox{info}}(\hat{E})=\ln (0.48 d)+\ln \sqrt{1-\zeta^2}
+\zeta^2\frac{(1-\hat{E}^2)}{2}\\\\
\Longrightarrow \exp S^{\mbox{info}}(\hat{E})=(0.48d)\,\sqrt{1-\zeta^2}\,
\exp\left\{\dis\frac{\zeta^2\,(1-\hat{E}^2)}{2}\right\}\;.
\earr
\label{eq.sinfo}
\ee
The minimum value of $S^{\mbox{info}}$ is $0$. While for GOE,
$S_{GOE}^{\mbox{info}}=\ln (0.48 d)$. Note that Eqs.(\ref{eq.npc}) and
(\ref{eq.sinfo}) are derived by assuming that strength fluctuations follow Porter-Thomas
(i.e. locally renormalized $C^E_k$ are Gaussian variables) and
several other assumptions as described in \cite{vkbrs01}. For embedded ensembles
the Porter-Thomas assumption and other assumptions are verified in some numerical
examples; see \cite{Brody,vkb2001,Ko-14} and references there in. The first order
corrections to Eqs.~(\ref{eq.npc}) and (\ref{eq.sinfo}) are also given in \cite{vkbrs01}. However, more
complete formulas for PR and $S^{\mbox{info}}$ for EGOE(1+2) and other embedded
ensembles are still not available.

Finally, here we will give the formula for $Q(\Omega,m,F) =
\overline{\sigma^2_{V(2)}(m,F)}$,
\begin{equation}
\barr{lcl}
Q(\Omega,m,F) &=& \dis\sum_{f=0,1}
(\Omega-1)(\Omega-2(-1)^f)(\Omega+2)\;P^{\nu=1,f}(m,F)\\\\
&+& \dis\frac{(\Omega-3)(\Omega^2+\Omega+2)}{2(\Omega-1)}
\; P^{\nu=2,f=0}(m,F)\\\\
&+& \dis\frac{(\Omega-1)(\Omega+2)}{2}\; P^{\nu=2,f=1}(m,F)
\earr
\label{eq.q1}
\end{equation}
where
\begin{equation}
\barr{l}
P^{\nu=1,f=0}(m,F) = \dis\frac{\l[(m+2)m^\star/2 - \lan F^2 \ran\r]
P^0(m,F)}
{8 (\Omega-2) (\Omega-1) \Omega (\Omega+1) } \;,\\\\
P^{\nu=1,f=1}(m,F) =\dis\frac{\l\{ \barr{l}  8\Omega(m-1)(\Omega+2m-4) \lan F^2 \ran \\
+(\Omega-2) P^2(m,F) P^1(m,F) \earr \r\}}
{8  (\Omega-1) \Omega (\Omega+1) (\Omega+2)^2} \;,\\\\
P^{\nu=2,f=0}(m,F) = \dis\frac{ \l[m^\star(m^\star-1) - \lan F^2 \ran\r]
P^0(m,F)}{8 \Omega (\Omega+1)}\;,\\\\
P^{\nu=2,f=1}(m,F) = \l\{ \l[\lan F^2 \ran\r]^2 (3\Omega^2+7\Omega+6)/2 +
3m(m-2) \times \r. \\ \l. m^\star (m^\star+1)
 (\Omega-1)(\Omega-2)/8 + \l[\lan F^2 \ran/2\r] \times  \r. \\ \l.
 \l[(5\Omega+3) (\Omega-2)m m^\star +
\Omega(\Omega-1)(\Omega+1)(\Omega-6)\r] \r\} \div  \\
 \l[ (\Omega-1) \Omega (\Omega+2)(\Omega+3) \r]\;; \\\\
P^0(m,F) = \l[ m(m+2) - 4F(F+1)\r]\;\;\;\;\\
P^1(m,F) = \l[3m(m-2) + 4F(F+1)\r]\;\;, \\
P^2(m,F) = 3(m-2) m^\star/2 + \lan F^2 \ran\;,\;\;\;\;\\m^\star =
\Omega+m/2\;,\;\;\;\; \lan F^2 \ran =F(F+1)\;.
\earr \label{eq.q2}
\end{equation}


\ed